\PassOptionsToPackage{unicode}{hyperref}
\PassOptionsToPackage{hyphens}{url}
\PassOptionsToPackage{dvipsnames,svgnames,x11names}{xcolor}
\documentclass[
  11pt,
]{article}
\usepackage{xcolor}
\usepackage[margin=1in]{geometry}
\usepackage{amsmath,amssymb}
\usepackage{iftex}
\ifPDFTeX
  \usepackage[T1]{fontenc}
  \usepackage[utf8]{inputenc}
  \usepackage{textcomp} 
\else 
  \usepackage{unicode-math} 
  \defaultfontfeatures{Scale=MatchLowercase}
  \defaultfontfeatures[\rmfamily]{Ligatures=TeX,Scale=1}
\fi
\usepackage{lmodern}
\ifPDFTeX\else
\fi
\IfFileExists{upquote.sty}{\usepackage{upquote}}{}
\IfFileExists{microtype.sty}{
  \usepackage[]{microtype}
  \UseMicrotypeSet[protrusion]{basicmath} 
}{}
\makeatletter
\@ifundefined{KOMAClassName}{
  \IfFileExists{parskip.sty}{%
    \usepackage{parskip}
  }{
    \setlength{\parindent}{0pt}
    \setlength{\parskip}{6pt plus 2pt minus 1pt}}
}{
  \KOMAoptions{parskip=half}}
\makeatother
\usepackage{longtable,booktabs,array}
\usepackage{calc} 
\usepackage{etoolbox}
\makeatletter
\patchcmd\longtable{\par}{\if@noskipsec\mbox{}\fi\par}{}{}
\makeatother
\IfFileExists{footnotehyper.sty}{\usepackage{footnotehyper}}{\usepackage{footnote}}
\makesavenoteenv{longtable}
\usepackage[]{natbib}
\usepackage{bookmark}
\IfFileExists{xurl.sty}{\usepackage{xurl}}{} 
\hypersetup{
  pdftitle={Vibe coding before the trend},
  pdfauthor={Leon van Bokhorst; Koen Suilen},
  colorlinks=true,
  linkcolor={blue},
  filecolor={Maroon},
  citecolor={Blue},
  urlcolor={blue},
  pdfcreator={LaTeX via pandoc}}

\title{Vibe coding before the trend}
\usepackage{etoolbox}
\makeatletter
\providecommand{\subtitle}[1]{
  \apptocmd{\@title}{\par {\large #1 \par}}{}{}
}
\makeatother
\subtitle{Practitioner Reflections on AI-Assisted Programming Across
Four Student Cohorts}
\author{Leon van Bokhorst \and Koen Suilen}
\date{March 2026}

\begin{document}
\maketitle

\vspace{-0.5em}
\begin{center}
\small
\begin{tabular}{@{}l@{:\quad}l@{}}
\textbf{Type} & Practitioner Report \\
\textbf{Organization} & Fontys ICT, University of Applied science \\
\end{tabular}
\end{center}
\vspace{0.75em}

\begin{center}\rule{0.5\linewidth}{0.5pt}\end{center}

\subsection{1. Why we are sharing this}\label{why-we-are-sharing-this}

During a session on AI with journalism students in our MIndlabs building
in Tilburg, something happened. It took a student a mere hour to build a
learning tool for dealing with misinformation. The student paused and
looked at us: We asked what's up and the student explained `this really
changes how I can study on this topic, fantastic!'. She had never worked
with code before and the idea of an IDE felt daunting the first minutes.

We are two senior lecturers at Fontys ICT, a University of Applied
Science in the Netherlands. We build, do, evaluate and learn. And when
something genuinely interesting happens, we usually tell each other
about it and that's often where it stops. That changed after visiting
the IFE conference in Monterrey \citep{ife2026}, where MIT's Justin
Reich strongly encouraged visitors to share findings on AI in education
even when experiments only feature small numbers or are not pre set-up,
because so little is known about the best approaches on learning with
AI.

In early 2025, Claude Sonnet 3.7 had just launched and the Cursor IDE
was gaining traction in developer communities. In our innovation
community of practice at Fontys ICT (iCoP), we were doing what educators
in fast-moving fields tend to do: debating what it meant for our
students. Leon picked up the thread: what if we just tried it? Give
students the tools, point them at a real problem or dataset to explore,
and see what happens?

What started as a two-day experiment with one cohort snowballed. A
second group. A visit to South Africa where the opportunity simply
presented itself. A journalism class that got an afternoon with tools
they'd never thought of using. In total: four settings, 107 student
reflections, and a slowly dawning realisation about what Andrej Karpathy
\citep{karpathy2025} would soon call ``vibe coding'': the idea that you
can create usable IT products by describing what you want, without ever
memorising a syntax rule.

We moved fast in a moment that felt urgent, document what we saw, and
then, months later, sit down to make sense of it together. The nature of
the changes we are dealing with and the speed with which they happen
call for an action based apraoch.

The world is not waiting. Every week, a new cohort of students is
encountering these tools for the first time, and the colleagues trying
to guide them through that don't have the luxury of holding off until a
cleaner study appears. We have four experiments, 107 student voices, and
a set of patterns worth sharing.

The question threading through this report: \emph{what happens when you
put frontier AI coding tools in front of students who weren't expecting
them, across different programmes, different backgrounds, different
countries and what can that tell us about where teaching and learning
might be heading?}

\begin{center}\rule{0.5\linewidth}{0.5pt}\end{center}

\subsection{2. Looking back: The playing field in
2025}\label{looking-back-the-playing-field-in-2025}

Early 2025 feels ages ago in terms of where we were on the AI maturity
ladder: Higher Educational institutes were still approaching AI use with
caution, developing policies and guidelines and almost no upfront
facilitation towards students. At Fontys ICT, we were drafting our AI
manifesto, an institutional guideline on how to approach generative AI
in learning.

We were taking the first steps towards a shared infrastructure and most
students and lecturers were using private accounts. From a technical
perspective, reasoning models such as o1 and o3 had just entered the
scene, Deepseek just made its entrance as it brought the Open Source
models to a new level and vibe coding still was in its very early
stages. In our Community of Practice (iCoP) we felt an urgency to
explore the impact of the capabilities of these models on democratizing
building IT products. What would happen if you give students a
challenge, a frontier AI model and let them discover possibilities?
Since we, the broad field of education, are in a phase where no one has
definitive answers, we are all working on hypothesis. So building,
exploring and learning are worth more than waiting for a proven
solution: learning by doing

During this period in time, we were already foreseeing the end of the IT
product as a proxy for learning so we had to find out in practice: How
do students approach the new possibilities opened by these, at the time,
frontier models and tools. Up until then, generated code was still seen
as unreliable, handled complexity was considered limited (limited
context windows) and practical use was often for many users still a
matter of copying code chunks instead of actually conversing about
solutions. This was the context in which we decided put our intuition to
the test.

\begin{center}\rule{0.5\linewidth}{0.5pt}\end{center}

\subsection{3. What we did: four
experiments}\label{what-we-did-four-experiments}

Leon came up with the initial concept of the Pokemon challenge: ``Create
an AI-powered analysis tool that answers the question: To evolve or not
to evolve?, determining when staying in current form might be better
than evolving. All the students got was a dataset from Kaggle, a early
version of the Cursor AI IDE and the question mentioned here. A group of
students was quickly found and the enthusiasm of the students was
inspiring, so we went looking for other groups to see what this
challenge could bring. Because we also sparked interest of other domains
and had an unexpected opportunity for a completely different cohort
during a conference trip, we could expand the experiment into other
contexts. What we noticed and obeserved, is the subject of this report.

In all settings, Cursor was used as the primary tool of choice. In some
occasions, students were also pointed towards more integrated solutions
like Lovable and Bolt. Student used a personal account and were limited
to the free tier (limiting the amount of AI requests made by those
tools)

{\def\LTcaptype{none} 
\begin{longtable}[]{@{}
  >{\raggedright\arraybackslash}p{(\linewidth - 10\tabcolsep) * \real{0.1455}}
  >{\raggedright\arraybackslash}p{(\linewidth - 10\tabcolsep) * \real{0.2000}}
  >{\raggedright\arraybackslash}p{(\linewidth - 10\tabcolsep) * \real{0.1091}}
  >{\raggedright\arraybackslash}p{(\linewidth - 10\tabcolsep) * \real{0.1818}}
  >{\raggedright\arraybackslash}p{(\linewidth - 10\tabcolsep) * \real{0.1455}}
  >{\raggedright\arraybackslash}p{(\linewidth - 10\tabcolsep) * \real{0.2182}}@{}}
\toprule\noalign{}
\begin{minipage}[b]{\linewidth}\raggedright
Cohort
\end{minipage} & \begin{minipage}[b]{\linewidth}\raggedright
Background
\end{minipage} & \begin{minipage}[b]{\linewidth}\raggedright
Size
\end{minipage} & \begin{minipage}[b]{\linewidth}\raggedright
Duration
\end{minipage} & \begin{minipage}[b]{\linewidth}\raggedright
Format
\end{minipage} & \begin{minipage}[b]{\linewidth}\raggedright
Reflection
\end{minipage} \\
\midrule\noalign{}
\endhead
\bottomrule\noalign{}
\endlastfoot
\textbf{Open Learning} & ICT (various) & 54 & 2 days & Teams of 6, build
+ review & Open essay \\
\textbf{Marketing} & Digital marketing minor & 24 & Half day &
Individual, open brief & Free-format text \\
\textbf{NWU (South Africa)} & BA Communication & 22 & Single session &
Individual, guided & Online form \\
\textbf{Journalism} & BA Journalism & 7 of 20 & Afternoon & Individual,
free diverge & Online form \\
\end{longtable}
}

\emph{Figure 1: Overview of the four cohorts and how each session was
set up.}

\subsubsection{3.1 Open Learning - The original pressure
cooker}\label{open-learning---the-original-pressure-cooker}

The Pokémon challenge originated here. A group of students that
participated in what Leon designed in collaboration with Claude 3.5 as
the 'great Pokémon evolution experiment''. It would give 54 Open
learning students a dataset, a question and a tool to help them achieve
something previously considered impossible in such a limited timeframe.
The 54 students with various ICT backgrounds (software, ux,
cybersecurity, business etc) worked in teams of 6 and had limited prior
AI/ML experience. Noteworthy for this group was the high degree of
engagement and little practical barriers due to the nature of their
background. Installing a library, reading code or using an unknown tool
was second nature for most of these students.

Day 1 focused on building, and students created various approaches to
tackle the question `to evolve or not?'. Creating Evolution decision
trees, Statistical pattern analysis, Power rankings and data
visualisations there was high degree of variation.

Day 2 focused on enhancement, code review and reflection: First students
had to analyse another team's solution, using Cursor for understanding
and codebase reviews. Secondly, they had to enhance the solution through
feature addition of performance optimalisation.

Finally, a structured reflection essay was part of the challenge, in
which this group especially mentioned aspects of career preparation and
impact on their learning approach.

\subsubsection{3.2 Marketing - Same idea, completely different
energy}\label{marketing---same-idea-completely-different-energy}

Koen knew this group, he already lectured on the subject of AI in
marketing. For the Digital marketing minor, a 3rd year semester mixing
analytics, marketing strategy and technology. This was a group of 24
students with a diverse background: ICT, Economics and Marketing being
the most prominent. Because Koen experienced from previous cohorts that
these students typically arrived with what you could call `IT anxiety',
this was an interesting group to see what they could achieve.

We adapted the challenge. The time window was half a day rather than two
because the path towards a feeling of success and the factors leading to
adoption or reluctance were what we were most interested in.
Additionally we therefore gave students more freedom: some built games,
some dove into data analysis. Afterwards, all 24 wrote free-format
reflections on what they'd experienced.

As one student here put it: ``Tools like Cursor allow us to look at
solutions from a functional perspective without coding prerequisites.
Learning becomes more interactive and exploratory rather than purely
documentation-based''

\subsubsection{3.3 NWU, South Africa - Same tools, different
world}\label{nwu-south-africa---same-tools-different-world}

During a conference visit as invited speaker, Koen had the chance to
engage with a class of third-year BA Communication students, enrolled in
a web-development course. While engaging with lecturers during the
conference and doing a campus visit, Koen asked if it was possible to do
an experiment with one of the themes discussed during the conference:
vibe coding. Chances like this don't present themselves everyday, so
while quickly improvising an idea on how to make this challenge work,
Koen managed to get an extra perspective on the gutfeel we started off
with.

One of the lecturers, Koos de Villiers was happy to hand over his class.
In this session, where students used the computer lab provided by North
West University, a similar challenge as the Digital Marketing students
was given and reflections were collected via online forms. The students
weren't used to this kind of open-ended challenge. Some were reluctant
at first, visibly intimidated by Cursor's interface. And like every
other group, they hit the usual early irritations: library dependencies,
unfamiliar install directories: the kind of issues that, being an IT
lecturer, you're not even aware of anymore. However, they can kill
enthusiasm and momentum before it starts.

These issues were common in all sessions, and once tackled, we were
surprised by how quickly the students adopted and just started creating
and discussing ideas, not tech. However, a snowball effect was visible
in the NWU context: the first succesfull student helped others and so
the majority of the attendees managed to get a working raw prototype.

\subsubsection{3.4 Journalism - Content over
technology}\label{journalism---content-over-technology}

This group was small in size, with only 7 students handing in a
reflection but 20 participating. However, they were remarkably
communicative on their experience. Both Leon and Koen hosted this
afternoon session with a group of students in their 2nd and 3rd year of
the BA program. Since they were doing a semester on `building and
creating in Journalism', they had a mindset open to discovering new ways
to achieve this.

While some more support on installation issues was needed, the use cases
here that students created when we gave them freedom if they want to
diverge from the inital Pokemon assignment were most interesting. As a
student here remarked along the lines of ``I wish I could have done this
in my first year, I could have used completely different learning
approaches''. Students here used the opportunity for example to create
personalized learning tools (a game, a self assessment) and focussed on
`close to home' use cases.

\begin{center}\rule{0.5\linewidth}{0.5pt}\end{center}

\subsection{4. What we saw: patterns in
reflections}\label{what-we-saw-patterns-in-reflections}

This challenge wasn't designed as a study, it started as a way to
collect stories and experiences from four very different contexts. Yet,
despite the obvious differences in those contexts, we did find some
emerging patterns when we debriefed after each session and read through
the 100+ reflections.

{\def\LTcaptype{none} 
\begin{longtable}[]{@{}
  >{\raggedright\arraybackslash}p{(\linewidth - 8\tabcolsep) * \real{0.3103}}
  >{\centering\arraybackslash}p{(\linewidth - 8\tabcolsep) * \real{0.1724}}
  >{\centering\arraybackslash}p{(\linewidth - 8\tabcolsep) * \real{0.1724}}
  >{\centering\arraybackslash}p{(\linewidth - 8\tabcolsep) * \real{0.1724}}
  >{\centering\arraybackslash}p{(\linewidth - 8\tabcolsep) * \real{0.1724}}@{}}
\toprule\noalign{}
\begin{minipage}[b]{\linewidth}\raggedright
Pattern
\end{minipage} & \begin{minipage}[b]{\linewidth}\centering
Open Learning
\end{minipage} & \begin{minipage}[b]{\linewidth}\centering
Marketing
\end{minipage} & \begin{minipage}[b]{\linewidth}\centering
NWU
\end{minipage} & \begin{minipage}[b]{\linewidth}\centering
Journalism
\end{minipage} \\
\midrule\noalign{}
\endhead
\bottomrule\noalign{}
\endlastfoot
\textbf{Career urgency} --- \emph{``If I don't learn this\ldots{}''} &
*** & *** & ** & * \\
\textbf{Syntax to thinking} --- \emph{AI handles the syntax} & *** & ***
& ** & ** \\
\textbf{From memorizing to evaluating} --- \emph{the skill shift} & ***
& ** & ** & * \\
\textbf{Partnership as principle} --- \emph{conductor, not replaced} &
*** & *** & *** & *** \\
\textbf{Accessibility} --- \emph{non-technical eyes} & * & ** & ** &
*** \\
\end{longtable}
}

\textbf{Legend:} \texttt{***} strongly present; \texttt{**} present;
\texttt{*} limited \emph{Figure 2: How the five patterns showed up
across the four cohorts.}

\subsubsection{4.1 If I don't learn this, I will fall
behind}\label{if-i-dont-learn-this-i-will-fall-behind}

The most consistent observation we felt that resonated with the
participants, was about career preparation. As a student noted: ``AI
tools like Cursor are reshaping software development. They don't replace
developers, but change our role to a focus on architecture and problem
solving. The ability to collaborate with AI is becoming increasingly
valuable.'' They view AI proficiency as an essential skill for their
future steps and development.

What struck us was how clear their observations were. There was very
little anxiety about replacement. Instead, there was a pragmatic
recognition that field had changed and they have to adapt.

\subsubsection{4.2 AI handles the syntax, students focus on
thinking}\label{ai-handles-the-syntax-students-focus-on-thinking}

A second pattern we came across, was how students view AI as a learning
enhancement partner, accelerating the process of problem solving and
being able to focus on a more high level view of problem solving:
``Working with Cursor has fundamentally changed my way of programming.
Instead of getting stuck on syntax errors, I can concentrate on core
concepts and system design.'' Open Learning students talked about
architecture and system design. Marketing students described being able
to focus on what they wanted to build rather than how to build it.

\subsubsection{4.3 From memorizing to
evaluating}\label{from-memorizing-to-evaluating}

A more nuanced pattern emerged around the question what skills actually
matter in this new reality. Students recognised that some traditional
skills such as memorising syntax were becoming less important. But they
were confident that critical thinking was becoming more important, not
less: ``Critical thinking and problem-solving skills are actually
becoming more important. You have to be able to properly assess whether
the suggestions are correct.'' Students were already buying into the
skill shift story: from producing to evaluating.

\subsubsection{4.4 Partnership as
principle}\label{partnership-as-principle}

Perhaps the most reassuring pattern for educators was the mental model
students used to describe their relationship with AI. Almost
universally, they framed it as collaboration. Words like ``partner,''
``assistant,'' and ``co-pilot'' appeared across all four cohorts.
Several students used the metaphor of a conductor directing an
orchestra: the AI provides the instruments, but the human sets the
direction. We found almost no expressions of fear about being replaced,
which contrasts sharply with the narrative in public media outings.

\subsubsection{4.5 Accessibility through non-technical
eyes}\label{accessibility-through-non-technical-eyes}

A final observation came from the non-IT participants. Particularly the
journalism and marketing cohorts didn't just appreciate the AI tools.
They described something more fundamental: a door opening to a domain
that had previously felt closed. A marketing student could now build an
IT product, using their own domain expertise as leverage rather than
being stopped by technical barriers. A journalism student wished they
had encountered these tools in their first year because it would have
changed their entire approach to learning.

What makes this observation stand out is that it appeared most strongly
in the groups with the least exposure time and the most constrained
reflection format. The journalism students had a single afternoon and a
structured survey form, yet accessibility was the theme they emphasised
most. That wasn't something we expected, and it raises a question we
find genuinely interesting: If the barrier to building IT solutions is
collapsing, what does that mean for how we think about who belongs in an
ICT or technology classroom?

\begin{center}\rule{0.5\linewidth}{0.5pt}\end{center}

\subsection{5. The bigger picture: then and
now}\label{the-bigger-picture-then-and-now}

In this chapter we reflect on the state of education and AI. A year on,
the landscape has shifted. Here's how what we saw in early 2025 relates
to where things are in spring 2026.

{\def\LTcaptype{none} 
\begin{longtable}[]{@{}
  >{\raggedright\arraybackslash}p{(\linewidth - 4\tabcolsep) * \real{0.3333}}
  >{\raggedright\arraybackslash}p{(\linewidth - 4\tabcolsep) * \real{0.3333}}
  >{\raggedright\arraybackslash}p{(\linewidth - 4\tabcolsep) * \real{0.3333}}@{}}
\toprule\noalign{}
\begin{minipage}[b]{\linewidth}\raggedright
\end{minipage} & \begin{minipage}[b]{\linewidth}\raggedright
\textbf{Early 2025} --- When we ran the experiments
\end{minipage} & \begin{minipage}[b]{\linewidth}\raggedright
\textbf{Early 2026} --- Writing this report
\end{minipage} \\
\midrule\noalign{}
\endhead
\bottomrule\noalign{}
\endlastfoot
\textbf{Tools} & Claude Sonnet 3.7 just launched. Cursor IDE gaining
traction. Minimal reasoning, no agentic capabilities. Limited context
windows. & Claude Code, OpenAI Codex, Google Antigravity. Agentic
workflows standard. Lovable and Bolt.new lower the entry barrier. \\
\textbf{Institutions} & HBO approaching AI with caution. Policies being
drafted. No shared infrastructure. Students on personal accounts. & AI
manifestos published. Some shared infrastructure emerging. Conversation
shifting from restriction to integration. \\
\textbf{Framing} & ``Vibe coding'' not yet a term. Code generation seen
as unreliable. Focus on cheating and detection. & Vibe coding widely
adopted. Students completing software development semesters without
writing code by hand. \\
\textbf{Student friction} & The barrier was plumbing, not prompting.
File structures, install paths, package dependencies. & Setup barriers
substantially lower. New questions: ownership, justification, judgement
of AI output. \\
\end{longtable}
}

\emph{Figure 3: How the landscape shifted between the experiments and
this report.}

\subsubsection{5.1 Then: what we assumed}\label{then-what-we-assumed}

We expected the AI tools, interacting, prompting and running the
solutions to be the hard part. We were wrong. When we started these
sessions in early 2025, Cursor's interface with black and white,
terminal-heavy, full of version control indicators looked intimidating,
and we assumed it would be. What we didn't anticipate was where the real
friction would sit. It wasn't in formulating prompts. Students were
surprisingly good at articulating what they wanted to build: a game, a
data analysis, a learning tool.

The bottleneck was far more basic than that. Students who had downloaded
Cursor couldn't find where the application had been installed on their
own computer. They opened projects in the wrong directory. They didn't
know what a file structure was. Before anyone could engage with
AI-assisted coding, they needed to navigate the unglamorous fundamentals
of how a computer organises files. This was further emphasized by the
tools' defaults.

If a student didn't specify a programming language, Cursor would
typically generate Python or Node.js solutions. Both of which come with
dependency management, virtual environments, and package installation
steps that are second nature to a developer but completely frustrating
to a journalism or marketing student. At the same time, a lack of basic
computer operations (file storage, folder management) also hindered more
students in the marketing and journalism cohorts than was foreseen and
required structured guidance from us. The biggest barrier to vibe
coding, it turned out, was not the ``vibe'' part. High on intent but low
on basic literacy, for many students the challenge was the plumbing and
the prompting.

Meanwhile, our colleagues at Fontys were largely indifferent. Vibe
coding wasn't on anyone's radar yet. The prevailing scepticism ran along
familiar lines: ``the code it generates isn't right,'' or ``it can't
handle real project complexity.'' There was limited institutional policy
which gave freedom to experiment, but no infrastructure to support it.
Students had to create their own accounts, use free personal licences.
We were essentially running these sessions on goodwill and curiosity.

\subsubsection{5.2 Now: what shifted}\label{now-what-shifted}

A little over a year later, quite a lot has changed. Some of it in ways
students predicted, some of it beyond anything we imagined. The syntax
observation has landed. We are now in a situation where a student can
complete a full semester of software development without writing a
single line of code by hand. That's not a thought experiment, it's our
current reality. What has shifted in response is where we place our
attention as educators. The questions that matter now are not ``can you
write this function'' but ``can you take ownership of what the AI
generated? Can you justify the decisions embedded in this code?''
Students in our 2025 reflections already pointed in this direction when
they wrote that syntax knowledge was becoming less important and that
they wanted to focus on ``the real problem.'' They were ahead of where
the curriculum was.

The tools have also moved dramatically. In early 2025, models had
minimal reasoning capability and no agentic features at all. The
landscape now includes Claude Code, OpenAI's Codex, Google's Antigravity
and a range of more accessible tools like Lovable and Bolt.new that
remove much of the interface friction we struggled with. Had we run the
same experiment today, the setup barriers would have been substantially
lower. This in itself raises an interesting question about how quickly
findings in this space become dated.

One thing students did not predict, because neither did we, was how
dramatically the tools would improve at understanding intent. In early
2025, it was still hard for students to think through a concept and then
have the AI build it. The gap between ``what I want'' and ``what the
tool produces'' required significant prompting skills. That gap has
narrowed considerably, which changes the nature of the skill shift we
observed, larger context windows and better overall context management.

\begin{center}\rule{0.5\linewidth}{0.5pt}\end{center}

\subsection{6. Back to the students:
signals}\label{back-to-the-students-signals}

We have observations rather than systemic follow-ups. But we did pick up
those signals, through end-of-semester assessments, informal
conversations, and feedback from fellow lecturers.

The clearest signal came from career choices. A notable number of
students from the marketing cohort went on to choose internships or
graduation projects where digital skills were central, because the
experiment had shown them that the boundary between ``technical'' and
``non-technical'' had shifted more than they'd assumed. The experience
didn't just change what they thought they could build. It changed what
they thought was within their reach professionally.

From the NWU cohort in South Africa, we got feedback that students
remembered the session as a significant moment. Though it's hard to
distinguish what made the impact: the tool, the open-ended format, or
the novelty of a visiting lecturer from abroad working in a
non-instruction-based way. Honestly, it was probably all three. We
cannot isolate factors but what we can say is that the session stood out
enough to be remembered and talked about months later, which is more
than most afternoon workshops achieve.

What we don't know is whether the experience persisted beyond the
initial enthusiasm. Did students continue using AI coding tools after
the session? Did their sense of possibility translate into sustained
practice? We suspect the answer varies enormously between individuals,
and we lack the data to say more than that.

\begin{center}\rule{0.5\linewidth}{0.5pt}\end{center}

\subsection{7. What we take from this}\label{what-we-take-from-this}

\subsubsection{7.1 For peers: what worked, what
didn't}\label{for-peers-what-worked-what-didnt}

If a colleague at another institution asked us for advice on trying
something similar, we'd say three things. First: just do it. The urge to
wait until you have the perfect setup, the right tools, institutional
approval, and a pre-registered research design will keep you waiting
while your students encounter these tools without any guidance at all.
Start messy. You'll learn more from an imperfect session than from a
perfect plan that never happens.

Second: don't underestimate interface friction. We started with Cursor,
which has a powerful but intimidating interface. For students without a
technical background, that interface was the barrier, not the AI itself.
If we did this again, we'd offer a range of tools with different levels
of complexity. For simple projects, chat-based interfaces like Claude's
artifacts or ChatGPT can get students to a result quickly. For more
ambitious work, tools like Lovable or Bolt.new provide a visual,
lower-friction entry point. IDE-based tools like Cursor are best
reserved for students who want more control over the process or are
working on complex, multi-component projects. We fell into the trap of
starting at the most complex end of the spectrum.

Third: let students bring their own problems. The Pokémon challenge
worked well for the Open Learning cohort because it was structured and
clearly defined. But for marketing and journalism students, Pokémon felt
irrelevant, they already had their own questions, their own domains,
their own use cases. When we gave them freedom to diverge, the quality
and motivation jumped. Next time, we'd offer two or three starting
points and explicitly invite students to bring their own challenge if
they have one.

One more thing we'd add: a pre- and post-measurement. Even something as
simple as three questions about how students view AI tools before the
session, repeated afterwards, would have given us a much stronger basis
for understanding what shifted. That's a small investment with
significant returns for anyone who wants to share their findings later.

\subsubsection{7.2 For ourselves: the methodological
lesson}\label{for-ourselves-the-methodological-lesson}

Moving along with the pace of AI and finding out what works and what
doesn't, asks for a pragmatic approach: action driven, agile and
reflective. That's how we approached this experiment.

Out of curiousity, we also looked at it from a more traditional research
perspective. The statistical analyses we ran on the 107 reflections
produced impressive numbers. For a moment, it looked like we had
discovered that educational background fundamentally shapes how students
experience AI tools.

Then we looked more carefully and realised we were mostly measuring our
own design choices. Students who had two days and an open document wrote
more than students who had an afternoon and a structured survey form.
The difference in richness was real, but it reflected our setup, not
their backgrounds.

The lesson is straightforward: when you analyse data that wasn't
collected for research purposes, the patterns you find may say more
about your collection method than about your participants.

\subsubsection{7.3 For the field: open
questions}\label{for-the-field-open-questions}

What we wanted to know is simple: what does this do to someone? To their
self-image, their view of their professional future, their sense of
what's possible? The qualitative observations and student quotes are
where the real value of this work sits. Not in statistical significance,
which was never the point, but in understanding how an experience with
AI tools reshapes a student's perspective on their own capabilities.

The questions we can't answer yet: Does the initial enthusiasm persist?
Does the ``skill shift'' students describe actually manifest in their
subsequent work, or does it fade once the novelty wears off? And perhaps
most importantly: as the tools become dramatically more capable and
accessible, do the patterns we observed in early 2025 still hold, or has
the landscape already shifted beneath them?

\begin{center}\rule{0.5\linewidth}{0.5pt}\end{center}

\subsection{8. An invitation}\label{an-invitation}

This report is addressed to fellow lecturers, educational researchers,
and curriculum designers in higher education, anyone who is navigating
the same questions we are about what AI means for how we teach and how
students learn.

We have four experiments, 107 student voices, a set of observations, and
the honest acknowledgement that we're figuring this out as we go.
Definitive answers are hard to find when the ground you are working on
moves as you walk. If that sounds familiar, we'd like to hear from you.
So here's a call to action: share your stories from the trenches,
through whichever medium or platform you might find suitable. We promise
to do the same.

To bring back Justin Reich's remarks at IFE 2026: turn on the lights and
talk about what you are doing, we cannot wait unitll someone else does
it for you.

\begin{center}\rule{0.5\linewidth}{0.5pt}\end{center}

\bibliography{references.bib}

\end{document}